\documentclass[11pt]{article}
\usepackage[margin=2.5cm]{geometry}
\usepackage{graphicx}
\usepackage{newtxtext}
\usepackage{newtxmath}
\usepackage{natbib}
\usepackage{hyperref}
\usepackage{soul}

\hypersetup{
    colorlinks = true,
    urlcolor   = blue,
    citecolor  = black,
}

\makeatletter
\long\def\@makecaption#1#2{%
  \vskip\abovecaptionskip
  {\small\leftskip\z@\rightskip\z@\parfillskip\z@ plus 1fil
   \parindent\z@ #1. #2\par}%
  \vskip\belowcaptionskip}
\makeatother

\newcommand{\backsection}[2][]{%
  \par\medskip\noindent
  {\small\textbf{#1.}\ #2}\par}
\usepackage{amsmath}

\newcommand\lrr[1]{\ensuremath{\left( #1 \right)}}

\usepackage[normalem]{ulem}
\usepackage{overpic}

\DeclareMathOperator{\sech}{sech}

\usepackage{tikz}

\usepackage{bm}

\title{Mode-2 solitary waves in continuous stratifications}

\usepackage{authblk}

\author[1]{Alex Doak\thanks{Corresponding author: a.doak@lboro.ac.uk}}
\author[1]{Ricardo Barros}
\author[2]{Paul A. Milewski}
\affil[1]{Department of Mathematical Sciences, Loughborough University, UK}
\affil[2]{Department of Mathematics, The Pennsylvania State University, State College, USA}

\date{\today}

\begin{document}
\maketitle

\begin{abstract}
Computed herein are examples of mode-2
internal solitary waves to the Dubreil-Jacotin--Long (DJL) equation.
These are localised, fully nonlinear waves in a continuously stratified, inviscid fluid under the Boussinesq and rigid-lid approximations. The equation is solved directly using finite differences on a periodic domain, enabling the computation of both generalised solitary waves, which possess oscillatory mode-1 tails, and embedded solitary waves,  whose tails vanish.
Two stratifications are considered. The first is a double pycnocline, generalising the three-layer system considered by \cite{doak2022large} by replacing each infinitesimal interface with a pycnocline of finite thickness.
The second stratification, a surface mixed layer above an exponentially stratified layer, shows that embedded mode-2 solitary waves are not restricted to double-pycnocline stratifications. These are the first mode-2 solitary waves of the DJL equation computed for stratifications that are asymmetric about mid-depth.
\end{abstract}

\section{Introduction}\label{sec:intro}

Internal waves are buoyancy-driven waves in stratified fluids. They are ubiquitous in the atmosphere and oceans, where they transport energy, mass, momentum and biological matter. Of particular interest are internal solitary waves (ISWs), localised, nonlinear waves of permanent form that can propagate over long distances with little change.

Continuously stratified fluids support infinitely many baroclinic modes of propagation, distinguished by their vertical structure.
Based on linear theory,
in a mode-1 wave, all isopycnals are displaced in the same direction, either upwards or downwards. In a mode-2 wave, some isopycnals are displaced upwards and others downwards.
Mode-1 ISWs have attracted considerably more attention than mode-2 ISWs, owing both to being more energetic \citep{shroyer2010mode} and easier to treat mathematically.
Mode-2 ISWs are nevertheless widely observed in the ocean \citep{yang2009observations, ramp2012observations} and in the laboratory \citep{honji1995experimental,carr2015experiments},
and may develop  trapped cores, which transport mass over long distances \citep{brandt2014laboratory, deepwell2016mass}.

The simplest stratification supporting mode-2 waves is a three-layer fluid, in which the density is a discontinuous piecewise-constant function \citep{lamb2000conjugate,
rusaas2002solitary,
kurkina2015propagation,
barros2020strongly}. This idealised stratification reflects the underlying issue with finding mode-2 ISWs in a general stratification, specifically that they typically travel at speeds within the linear spectrum of mode-1 waves. As a consequence, they usually occur as generalised solitary waves (GSWs), carrying a mode-1 oscillatory tail \citep{akylas1992solitary}. \cite{doak2022large} found particular parameters of the three-layer model for which these tails vanish, resulting in so-called embedded solitary waves (ESWs), in that they are \emph{embedded} in the continuous spectrum \citep{yang1999embedded, champneys2001embedded}. They form a codimension-one subset of the family of GSWs, so finding them requires a search through the GSW solution space.
A complicated bifurcation structure was discovered, including multihump ESWs, solitary waves with multiple oscillations about the central pulse. As well as ESWs, it was found that mode-2 solitary waves could have speeds greater than that of the mode-1 long-wave speed. In that model the flow in each layer is irrotational, so vorticity is confined to the interfaces. In this paper we show that these features survive when the interfaces are replaced by pycnoclines of finite thickness, as in real stratifications.

In a continuous stratification, steady solutions of the Euler equations satisfy the Dubreil-Jacotin--Long (DJL) equation \citep{DJ, long1953some}. Many computations of its solitary waves use the method of \cite{turkington1991computational}, in which solitary waves are minimisers of the kinetic energy at fixed available potential energy. A spectral implementation of this approach by \cite{dunphy2011spectral}, the DJLES code, is now widely used (for a review, see \citeauthor{stastna2022internal} \citeyear{stastna2022internal}).  However, \cite{turkington1991computational} show that minimisers have displacements of one sign, and hence mode-1 vertical structure; mode-2 and higher-mode waves are saddle points of the constrained functional, not minimisers. They cannot be obtained in this way unless the Boussinesq approximation is adopted and the waves are reflectively symmetric about mid-depth. In this case, they are equivalent to mode-1 waves in a channel of half the depth, reflected about the mid-depth streamline. To date, these are the only mode-2 solutions of the DJL equation that have been computed and the only class for which existence has been rigorously established \citep{tung1982large, zhang2026numerical}.

In this work we instead discretise the DJL equation directly with finite differences, following \cite{davis1967solitary} and \cite{tung1982large}. Because this approach does not rely on a variational characterisation, it places no restriction on the vertical structure. We use it to compute mode-2 solitary waves in stratifications that are not symmetric about mid-depth. We consider two stratifications: a double pycnocline, and a surface mixed layer above an exponentially stratified layer.
\section{Formulation and numerical method}\label{sec:formulation}
Consider an incompressible, inviscid fluid bounded above and below by rigid walls a distance $H$ apart. We take Cartesian coordinates $(x,z)$ with the origin on the lower wall, so that the rigid lid is at $z=H$.
We seek waves of permanent form propagating horizontally with constant speed $c$, for which the flow is steady in a moving frame. Introducing the streamfunction $\psi(x,z)$ with velocity $(u,w)=(\psi_z,-\psi_x)$, conservation of mass implies that the density is constant along streamlines, so that $\rho=F(\psi)$.
Assuming that every streamline originates far upstream, where the flow is a uniform stream with undisturbed density,
\begin{align}\label{eq:farfield}
\psi\rightarrow cz, && \rho\rightarrow\rho_0\bar\rho(z) && \text{as } x\rightarrow\pm\infty,
\end{align}
which gives $F(\psi)=\rho_0\bar\rho(\psi/c)$.
Under the Boussinesq approximation, in which density variations are retained only in the buoyancy term, conservation of momentum reduces to the DJL equation
\begin{align}\label{eq:DJL_psi}
\nabla^2 \psi =-\frac{g}{\rho_0}\lrr{z - \frac{\psi}{c}} F'(\psi), && \psi(x,0)=\psi(x,H)-cH=0.
\end{align}
This is a vorticity equation in which the vorticity is generated solely by baroclinic effects. In terms of the streamline displacement $\eta=z-\psi/c$, it reads
\begin{align}
\nabla^2 \eta + \frac{N^2(z-\eta)}{c^2}\,\eta = 0, && \eta(x,0)=\eta(x,H)=0,
\end{align}
where $N^2=-g \,\mathrm{d}\bar\rho/\mathrm{d}z$ is the squared buoyancy frequency of the undisturbed state, and the density is $\rho(x,z)=\rho_0\bar\rho(z-\eta(x,z))$.
Linearising about $\eta=0$ and considering long waves by neglecting $\eta_{xx}$ gives the eigenvalue problem for the vertical structure function $\phi(z)$,
\begin{align}\label{eq:longwave}
\phi'' + \frac{N^2(z)}{c^2}\phi = 0, && \phi(0)=\phi(H)=0,
\end{align}
whose eigenvalues $c_0^{(1)}>c_0^{(2)}>\dots$ are the linear long-wave speeds of modes $1,2,\dots$; the mode-$n$ eigenfunction has $n-1$ interior zeros. Linear mode-1 waves of finite wavelength travel with speeds in the interval $(0,c_0^{(1)})$, so a mode-2 solitary wave with $c<c_0^{(1)}$ lies within the mode-1 continuous spectrum and can resonate with them.

Much numerical work on the DJL equation uses the seminal method devised by \cite{turkington1991computational}. In their iterative scheme a solitary wave is characterised as a minimiser of the kinetic energy $KE$ subject to a fixed available potential energy $APE$, where
\begin{align}
    KE &= \frac{\rho_0 c^2}{2}\iint_D \,\lvert\nabla\eta\rvert^2 \,\mathrm{d}x\,\mathrm{d}z, \label{eq:KE}\\
    APE &= \rho_0\iint_D \int_0^{\eta} \xi\, N^2(z-\xi)\,\mathrm{d}\xi\,\mathrm{d}x\,\mathrm{d}z. \label{eq:APE}
\end{align}
The wave speed then appears as the Lagrange multiplier associated with the $APE$ constraint. \cite{turkington1991computational} show that, for any solution symmetric in $x$, minimisers can be taken to be of one sign, and hence have mode-1 vertical structure; higher-mode waves cannot be obtained variationally unless they are reflectively symmetric about mid-depth. Symmetric mode-2 Boussinesq solutions with symmetric stratification about the mid-depth are amenable to the variational approach as they are equivalent to mode-1 solutions on half the domain.

\cite{doak2022large} computed general mode-2 ESWs for the three-layer model. We suspected, and show in the following section, that these solutions exist for the case of continuous stratifications limiting to the layered model as pycnocline thickness goes to zero. Given that the method of \cite{turkington1991computational} cannot recover these solutions, we turn to more classical finite difference approaches used by \cite{davis1967solitary} and \cite{tung1982large}. Equation \eqref{eq:DJL_psi} is solved numerically on a rectangular domain in $(x,z)$ via a finite difference scheme.  Second-order central differences and periodic boundary conditions in $x$ are used. Many mode-2 solitary waves are in fact GSWs with periodic tails, and cases where the tail has zero amplitude (ESWs) are a subset of solutions. The domain is taken sufficiently long that for a tailless solitary wave, further increasing the domain length has negligible effect on the solution.
The numerical method  closely follows that of \cite{doak2025stokes} to compute free-surface waves with non-constant vorticity.

As discussed at length by \cite{doak2022large}, computing an ESW can be difficult, since ESWs do not bifurcate from the linear long-wave speed, but occur as isolated points on branches of GSWs. Once an ESW is computed, further ESWs can be found by continuation. Since ESWs form a codimension-one subset of the GSW family, an additional constraint is necessary. We set the curvature (denoted $K\approx \psi_{xx}$) of the streamline at mid-depth at the periodic edge (where $\psi_x=0$) of the computational domain  to zero, ensuring the solution is tailless \citep{champneys2001embedded}. In doing so, an additional unknown must be introduced, for which we take a parameter of the stratification: the lower-layer thickness $H_3$ for the double pycnocline (as in \citeauthor{doak2022large}) and the mixed-layer depth $z_0$ for the surface mixed-layer stratification (\S\,\ref{sec:results}).
\subsection{Trapped cores}\label{sec:cores}
 Equation \eqref{eq:DJL_psi} relies on the far-field condition \eqref{eq:farfield}, and therefore assumes that every streamline originates at $x\rightarrow-\infty$ and terminates at $x\rightarrow\infty$. However, this condition is unnecessarily restrictive \citep{brown1998fully}, and
 solutions of the DJL equation have been found that develop regions of closed streamlines, forming a trapped core \citep{tung1982large}. Streamlines within the core never reach the far field, so the upstream state places no constraint on the density or vorticity there, and the continuation of \eqref{eq:DJL_psi} into the core is a free choice. Solving \eqref{eq:DJL_psi} throughout the domain corresponds to the simplest continuation, which is to extend the far-field relations into the core. The resulting core is statically unstable, with denser fluid overlying lighter fluid. More physically motivated core models have been proposed \citep{helfrich2010model, carr2012instability}. However, the resulting waves were still found to be unstable when integrated in time, and the correct form of the flow within a trapped core remains an open problem. These models also require the value of the streamfunction on the core-bounding streamline. For mode-1 waves this value is readily obtained, since the core must lie against a boundary \citep{tung1982large}. For mode-2 waves, and in particular those that are not symmetric about the mid-depth, the core may lie in the interior of the domain, and identifying its bounding streamline is non-trivial. Since the aim of the present work is to establish the existence and structure of mode-2 solitary waves, we do not pursue these complications here. Solutions with trapped cores should therefore be understood as being unstably stratified within the core. We also note that many of the solutions presented here do not possess trapped cores.
\section{Results}\label{sec:results}
\begin{figure}
    \centering
     \begin{overpic}[scale=1]{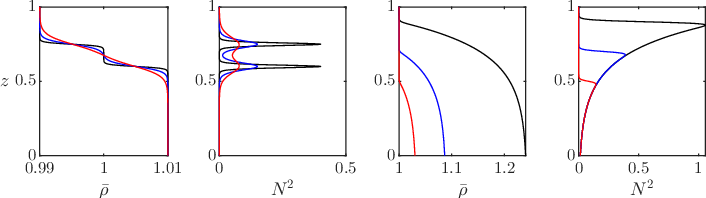}
     \put(15,30){\textbf{Double pycnocline} }
     \put(65,30){\textbf{Surface mixed layer} }
          \put(-1,25){$(a)$}
          \put(24,25){$(b)$}
          \put(50,25){$(c)$}
          \put(75.5,25){$(d)$}
     \end{overpic}
    \caption{Stratification profiles considered in this paper. Panels $(a)$ and $(b)$ show the density and $N^2(z)$ profiles of the double-pycnocline stratification \eqref{eq:strat_dp} with $H_3=0.6$ and $H_2=0.15$, where the black, blue and red curves correspond to $\alpha=80, 30$ and $15$, respectively. Panels $(c)$ and $(d)$ show the density and $N^2(z)$ profiles of the surface mixed-layer stratification \eqref{eq:strat_ml}, with the black, blue and red curves corresponding to $z_0=0.1, 0.3$ and $0.5$, respectively. \label{fig:strat}}
\end{figure}

We now use the numerical method of \S\,\ref{sec:formulation} to compute mode-2 solitary waves for two stratifications. We first consider a double-pycnocline stratification, for which recent work \citep{barros2020strongly,doak2022large} on the three-layer model with homogeneous layers provides guidance on where to find solutions. We then present results for a surface mixed layer above an exponentially stratified layer, typical of the upper ocean. Throughout, we fix $g=1$, $H=1$, and $\rho_0=1$, so that lengths are scaled by $H$ and speeds by $\sqrt{gH}$. Figure \ref{fig:strat} shows typical stratifications for the two cases.

\subsection{Double-pycnocline stratification}
We begin with a stratification profile of the form
\begin{align}\label{eq:strat_dp}
N^2(z) &= \frac{\alpha\Delta}{2}\sech^2\lrr{\alpha \lrr{z-H_3}}   + \frac{\alpha\Delta}{2}\sech^2\lrr{\alpha \lrr{z-H_3-H_2}}.
\end{align}
This represents a smoothed three-layer configuration with lower-layer thickness $H_3$ and middle-layer thickness $H_2$, the layers being numbered from the top. The upper-layer thickness is then $H_1=1-H_2-H_3$. The parameter $\Delta=0.01$ is the density jump across each pycnocline, and $\alpha$ is the inverse width of the pycnoclines. Figure \ref{fig:strat}$(a)$,$(b)$ shows the stratification for $H_3=0.6$, $H_2=0.15$ and varying $\alpha$. In the limit $\alpha\rightarrow\infty$, we recover three homogeneous layers separated by infinitesimal interfaces. As $\alpha$ decreases, the pycnoclines broaden.

\begin{figure}
    \centering
    \begin{overpic}[width=14cm]{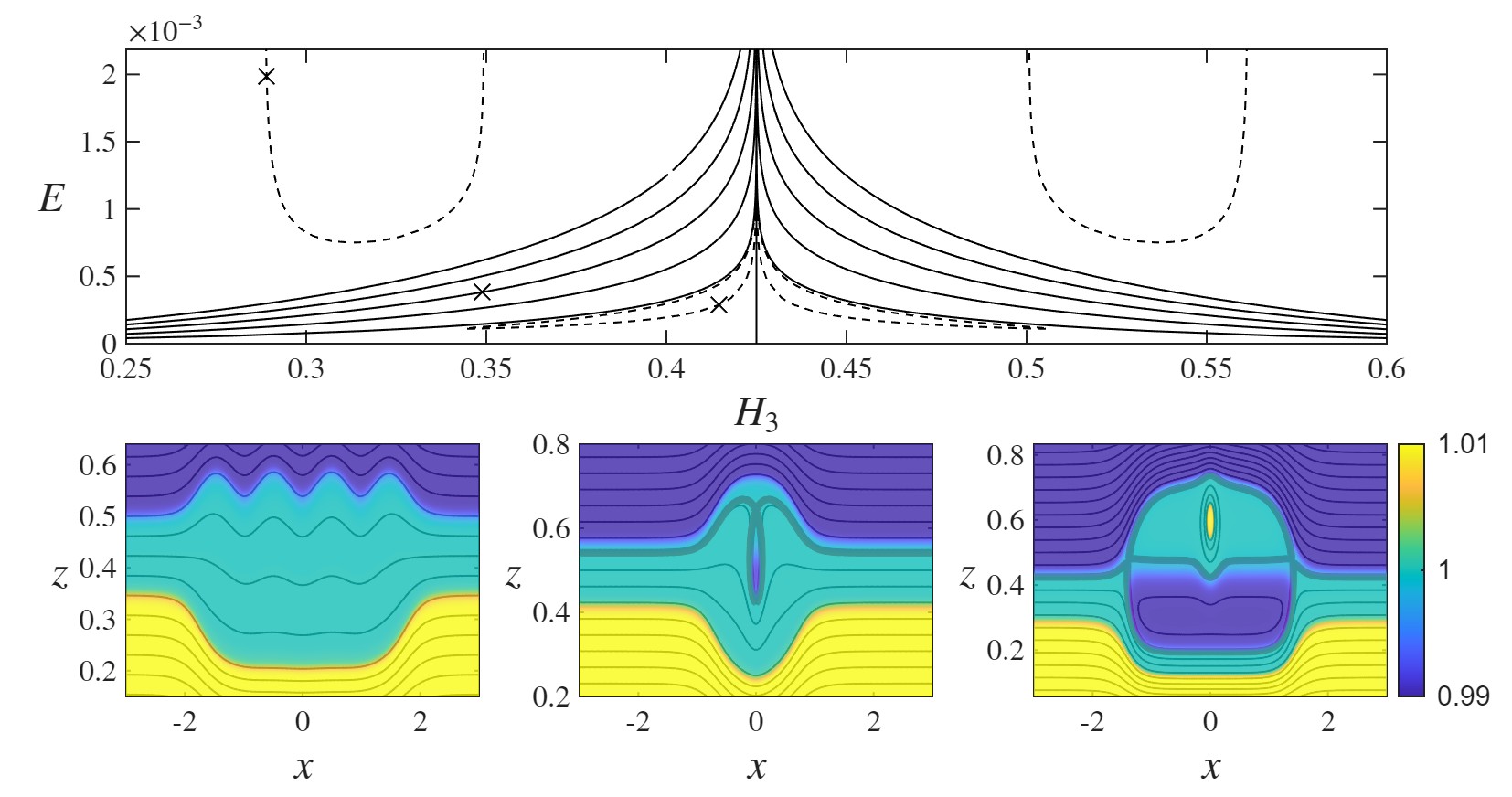}
    \put(1,22){$(a)$}
    \put(31.5,22){$(b)$}
    \put(61.5,22){$(c)$}

    \put(38,45){$(a)$}
    \put(47,31){$(b)$}
    \put(18,47){$(c)$}

     \put(33,34){%
        \begin{tikzpicture}
            \draw[->, line width=0.8pt]
                (0,0) -- (-.7,-1.4);
        \end{tikzpicture}%
    }
    \end{overpic}
    \caption{ESW branches for the double pycnocline \eqref{eq:strat_dp} with $H_2=0.15$ and $\alpha=80$. The upper panel shows the branches in the $(H_3,E)$-space. Solid (dashed) curves denote solutions without (with) a trapped core. Three profiles are shown in panels $(a)$--$(c)$. In addition to the multihump ESW solutions also found in the three-layer model (panel $(a)$), we identify solutions with trapped cores (panels $(b)$,$(c)$). The streamlines bounding the core are shown in bold, and the colormap corresponds to density. \label{fig:dp_1}}
\end{figure}
We consider two values of the middle-layer thickness $H_2$, each held fixed. In the first case, we take $H_2=0.15$. In the three-layer limit $\alpha\rightarrow\infty$, the mode-2 solitary waves are convex (the upper (lower) pycnocline has positive (negative) isopycnal displacement) and travel more slowly than the mode-1 long-wave speed $c_0^{(1)}$, so that they lie within the mode-1 continuous spectrum and only isolated tailless solutions (ESWs) exist. In the second case, we consider $H_2=0.8$, for which the three-layer limit supports concave mode-2 solitary waves (the upper pycnocline is displaced downwards and the lower upwards). These include both ESWs, with $c<c_0^{(1)}$, and regular ({\it i.e.,} outside the linear spectrum) mode-2 ISWs with $c>c_0^{(1)}$.

\subsubsection{Convex waves, $H_2=0.15$}
We begin by taking a thin pycnocline ($\alpha=80$) and comparing the solution space with that of the layered model. Rather than sweep through GSW branches to find an ESW, we construct a good initial guess for a mode-2 ESW by modifying the solution in figure 18 of \cite{doak2022large}. We reconstruct the streamfunction of the three-layer wave throughout the fluid and smooth it across the interfaces, where the velocity is discontinuous. By adjusting the smoothing width by trial and error, our DJL solver can be made to converge to a mode-2 ESW.
 \begin{figure}
    \centering
    \begin{overpic}[width=14cm]{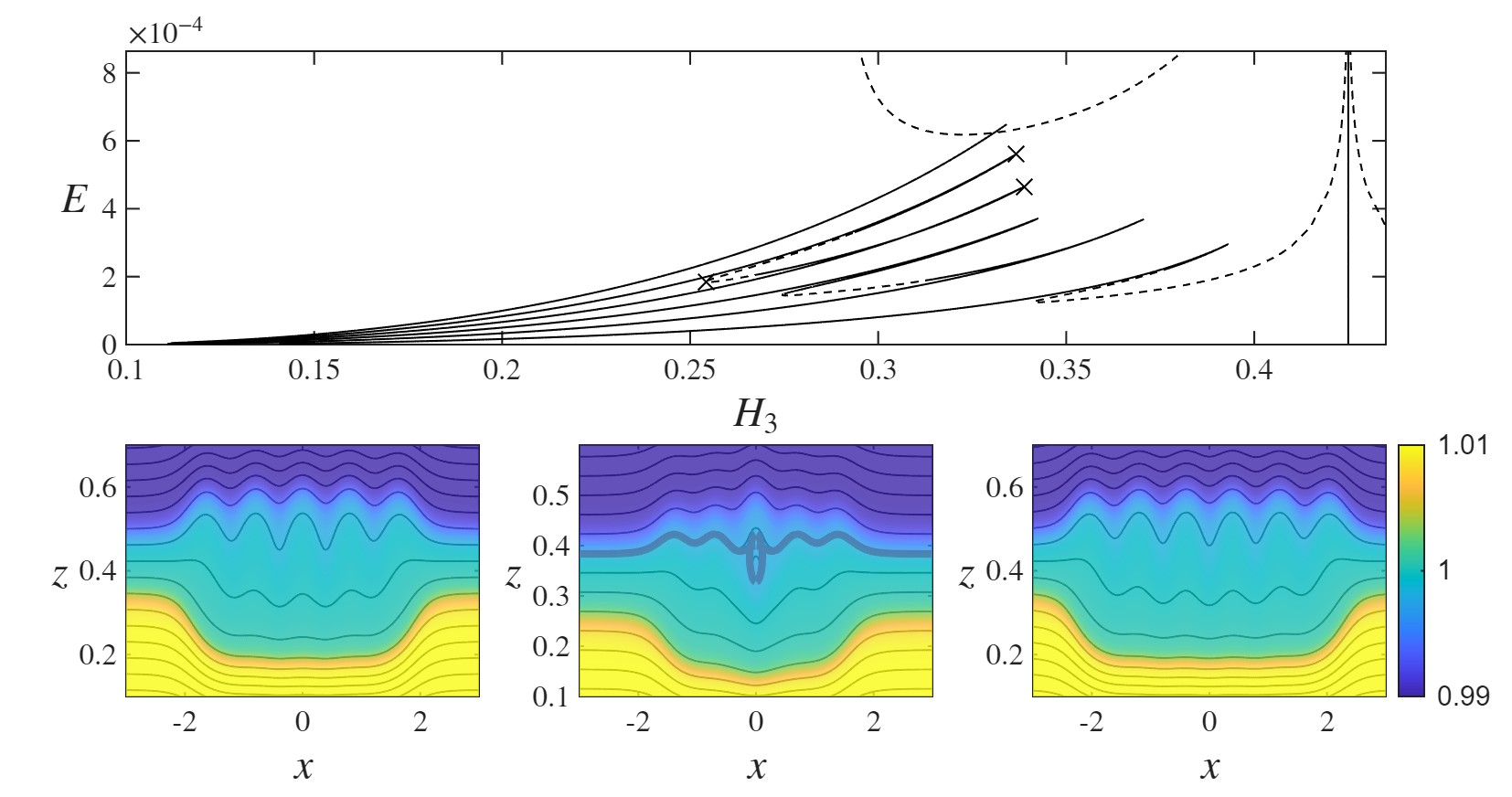}
    \put(1,22){$(a)$}
    \put(32,22){$(b)$}
    \put(62,22){$(c)$}

    \put(68.5,40){$(a)$}
    \put(67.5,42.5){$(c)$}
    \put(33,39){$(b)$}

     \put(36,34){%
        \begin{tikzpicture}
            \draw[->, line width=0.8pt]
                (0,0) -- (1.4,-0.7);
        \end{tikzpicture}%
    }
    \end{overpic}
    \caption{Same as in figure \ref{fig:dp_1} but with $\alpha=30$. We show the solution branches for $H_3\leq H_1$; those with $H_3>H_1$ are recovered by reflection about $H_3=0.425$. The multihump branches, which are disconnected for $\alpha=80$, now connect to one another. As one moves along the branch from solution $(a)$ to solution $(c)$, an additional oscillation in the central pulse is obtained.
    \label{fig:dp_2}}
\end{figure}

Figure \ref{fig:dp_1} shows the solution space for $\alpha=80$. The upper panel shows solution branches in $(H_3,E)$-space, where $E=KE+APE$ is the total energy. Solid curves denote solutions without a trapped core (see \S\,\ref{sec:cores}), while dashed curves correspond to solutions with a trapped core. A branch of solutions that are symmetric about mid-depth lies on the line $H_3=0.425$, for which the upper and lower layers have equal thickness. On either side of this branch lies a sequence of multihump solution branches (see solution $(a)$). As $E$ increases, these branches approach $H_3=0.425$ asymptotically and the solutions broaden into a tabletop solitary wave with additional oscillations superposed on the conjugate state \citep{lamb2000conjugate}. As $E$ decreases, the pycnoclines move towards the upper or lower wall, and the amplitude of the waves becomes very small. It remains unclear how the branches end, as the method ceases to converge in this region. The solution structure described above is qualitatively the same as in the limit $\alpha\rightarrow\infty$ \citep{doak2022large}. These solutions do not have trapped cores. We also find two additional branches, along which the solutions have trapped cores. Solution $(b)$ lies on one of these branches. Interestingly, the bounding streamline of the core does not have two stagnation points, one on either side of the core, but a single stagnation point above it. To our knowledge, a core bounded in this way has not been reported before. It cannot occur for a mode-1 ISW in a fluid at rest, whose core lies against a wall and is bounded by a streamline with two stagnation points on that wall \citep{tung1982large}. As $E$ increases along these two branches, the waves broaden (see solution $(c)$).

\begin{figure}
    \centering
    \begin{overpic}[width=14cm]{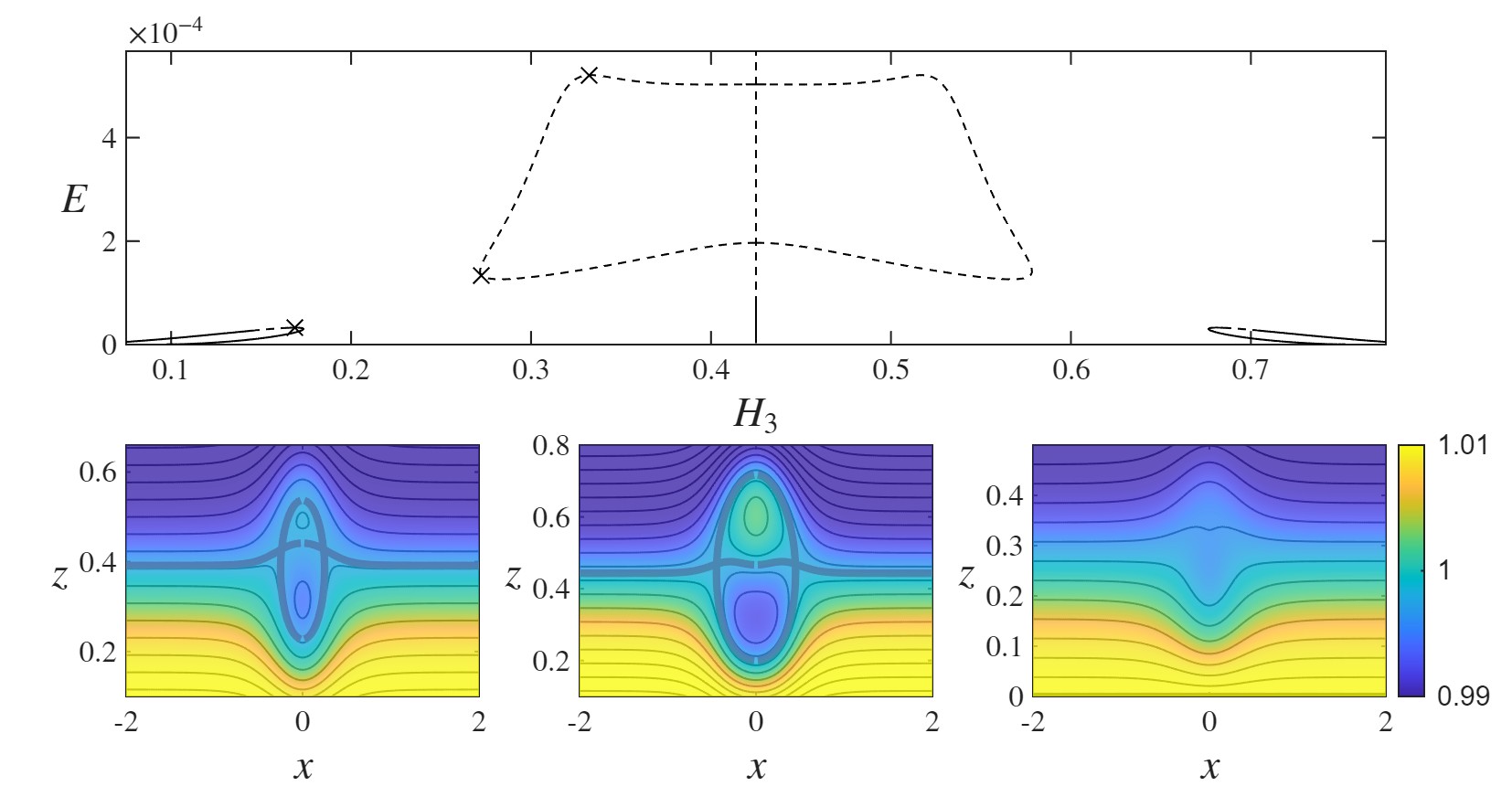}
    \put(1,22){$(a)$}
    \put(32,22){$(b)$}
    \put(62,22){$(c)$}

    \put(28,33){$(a)$}
    \put(35,47.5){$(b)$}
    \put(18,32){$(c)$}
    \end{overpic}
    \caption{Same as in figure \ref{fig:dp_1} but with $\alpha=15$. Two symmetry-breaking bifurcations occur along the symmetric $H_3=0.425$ branch. Most solutions have trapped cores, such as those shown in panels $(a)$ and $(b)$.  Panel $(c)$ shows a solution without a trapped core.  \label{fig:dp_3}}
\end{figure}

Figure \ref{fig:dp_2} shows the solutions computed for $\alpha=30$. The upper panel shows only half the $(H_3,E)$-space; the other half follows by reflection about $H_3=0.425$, a consequence of the up--down symmetry of the Boussinesq problem. The multihump branches, which for $\alpha=80$ approach the symmetric branch asymptotically, instead connect to each other. Furthermore, some solutions along these branches, like solution $(b)$, have trapped cores. As $\alpha$ is decreased further, the solver fails to converge when $K=0$ is imposed along most of these multihump branches. All the ESWs found at $\alpha=15$ are shown in figure \ref{fig:dp_3}. A symmetric branch again lies on $H_3=0.425$. Along this branch are two symmetry-breaking bifurcation points, and the asymmetric branch that connects them forms a loop. Solutions along this loop all have the typical single-hump (bulge) shape (see panels $(a)$,$(b)$). An additional solution branch of small amplitude was found, whose profiles appear to have a (2,1) multihump structure (two oscillations on the upper pycnocline, one oscillation on the lower), as seen in solution $(c)$. We were unable to continue the other multihump branches to $\alpha=15$, although this does not rule out their existence.
  \begin{figure}
    \centering
    \begin{overpic}[width=14cm]{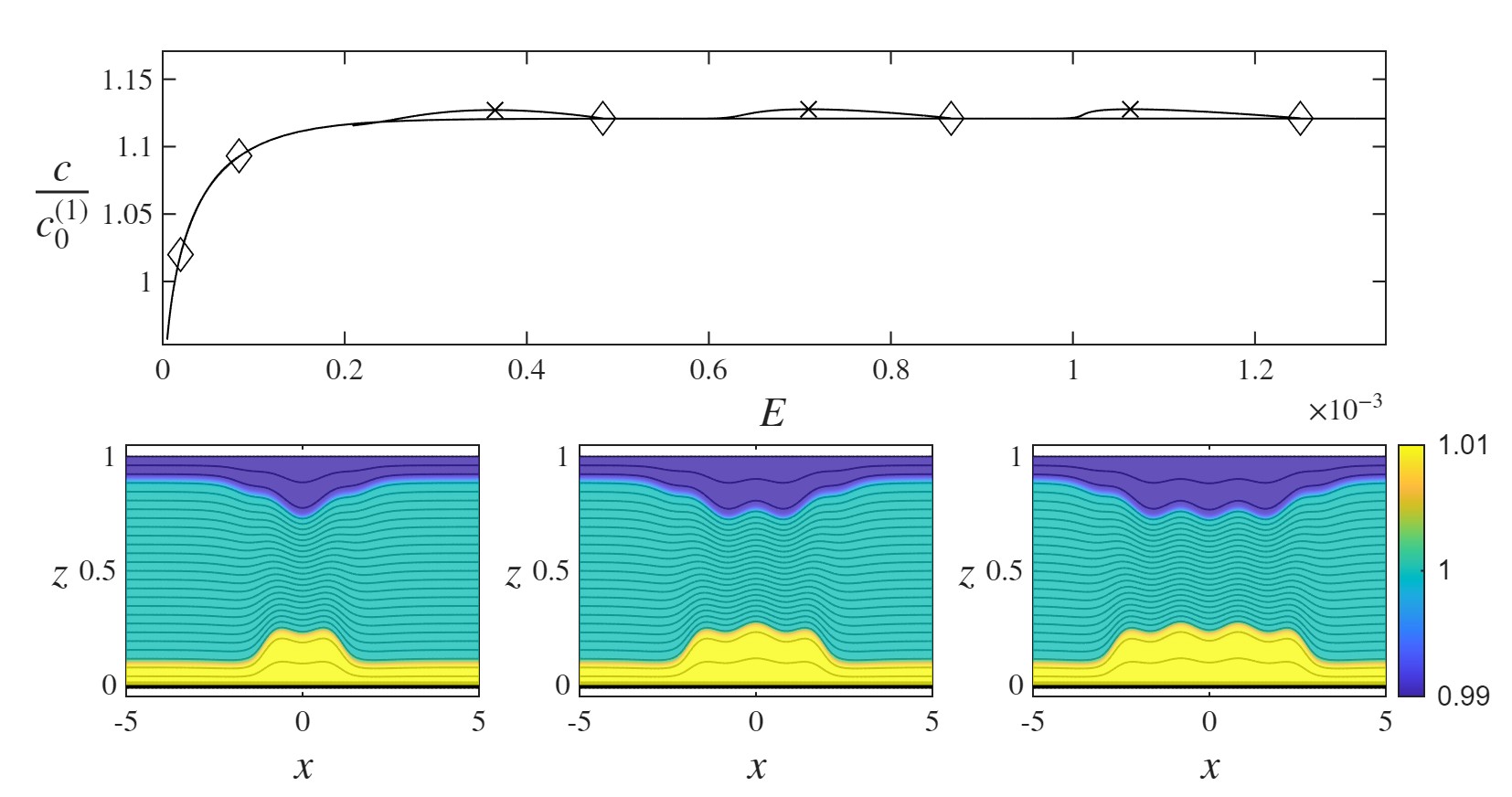}
    \put(1,22){$(a)$}
    \put(32,22){$(b)$}
    \put(62,22){$(c)$}

    \put(31,47){$(a)$}
    \put(52,47){$(b)$}
    \put(73,47){$(c)$}
    \end{overpic}
    \caption{Solitary wave branches for the concave case with $H_3=0.1$, $H_2=0.8$, and $\alpha=80$. The curve that bifurcates from the mode-2 long-wave speed and approaches a limiting speed as the wave broadens is the symmetric solution branch, on which the streamline at mid-depth is horizontal. Along the branch is a sequence of symmetry-breaking bifurcations, marked by diamonds. The solutions $(a)$--$(c)$ show that each symmetry-breaking branch adds another oscillation to the broadened section.
    \label{fig:concave_s2=80} }
\end{figure}

\subsubsection{Concave waves, $H_2=0.8$}
The second case we consider is $H_2=0.8$. The upper panel of figure \ref{fig:concave_s2=80} shows the solution space for $H_3=0.1$, $H_2=0.8$ and $\alpha=80$ in the $(c/c_0^{(1)},E)$-space. This stratification is symmetric about mid-depth, and a symmetric solution branch bifurcates from the mode-2 long-wave speed and approaches a broadened tabletop solitary wave with speed $c_m<c_0^{(1)}$. This branch represents solutions whose mid-depth streamline is flat. They are mode-1 solitary waves in a channel of half the depth, reflected about mid-depth (\S~\ref{sec:intro}), and thus have no mode-1 tail. As was found for the three-layer model, a number of symmetry-breaking bifurcations, shown as diamonds, occur along this branch. These bifurcations, and the asymmetric branches emanating from them, occur at speeds $c>c_0^{(1)}$, and hence the asymmetric solutions are not ESWs but form a continuous branch of mode-2 ISWs for a fixed stratification profile. The solution structure again shows multihump behaviour, with each successive branch adding an extra oscillation to the broadened wave, as shown by solutions $(a)$--$(c)$.

\subsection{Surface mixed layer}
Consider a stratification consisting of a surface mixed layer above an exponentially stratified layer, as used by \cite{lacasce2015estimating},
\begin{align}\label{eq:strat_ml}
N^2(z)  &=  N_0^2\exp\lrr{\frac{z-H}{e_{\text{fold}}}} \lrr{1-\tanh\lrr{\alpha(z-H+z_0)}},
\end{align}
where $N_0^2=1$ is the squared buoyancy frequency scale, $e_{\text{fold}}=0.2$ is the e-folding depth, $\alpha=80$ is the inverse thickness of the transition between the layers, and $z_0$ is the depth of the mixed layer. We fix all the stratification parameters except $z_0$, which we vary when computing ESW branches. The stratification is shown in panels $(c)$,$(d)$ of figure \ref{fig:strat}.
 \begin{figure}
    \centering
    \begin{overpic}[width=14cm]{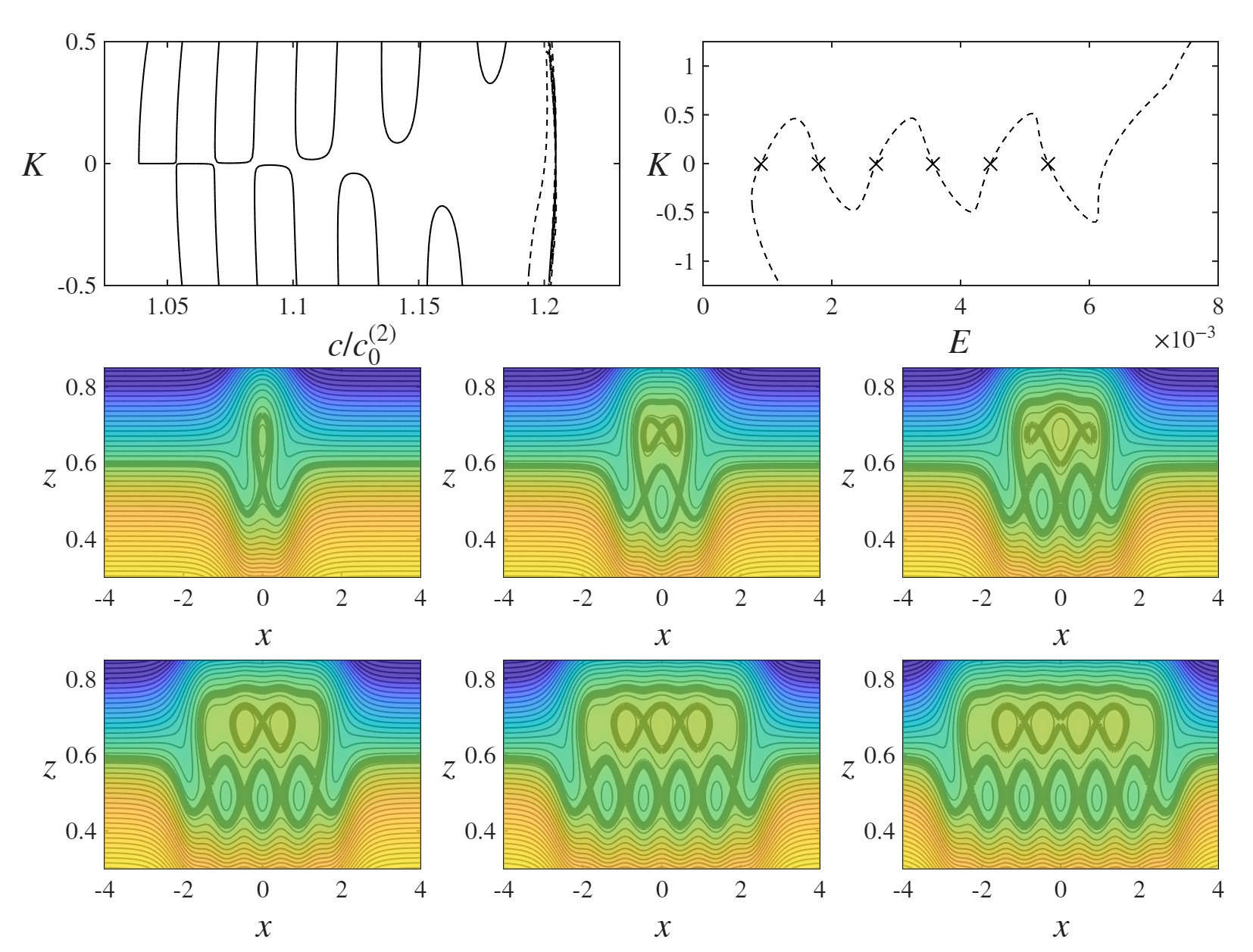}
    \put(1,45){$(a)$}
    \put(33.5,45){$(b)$}
    \put(65.5,45){$(c)$}

    \put(1,22){$(d)$}
    \put(33.5,22){$(e)$}
    \put(65.5,22){$(f)$}

    \put(58,63.5){\tiny $(a)$}
    \put(67.5,63.5){\tiny $(c)$}
    \put(77,63.5){\tiny $(e)$}

    \put(62.5,62){\tiny $(b)$}
    \put(72,62){\tiny $(d)$}
    \put(81,62){\tiny $(f)$}

    \end{overpic}
    \caption{GSW branches for the stratification \eqref{eq:strat_ml} with $z_0=0.2$. The upper-left panel shows a sequence of n- and u-shaped branches of GSWs in the $(c/c_0^{(2)},K)$-space. At $c\approx 1.2c_0^{(2)}$, a branch that repeatedly passes through $K=0$ is found. This branch is plotted in the $(E,K)$-space in the upper-right panel. Several ESWs are shown in panels $(a)$--$(f)$. The solutions all have trapped cores, with remarkably complex structure not observed for mode-1 ISWs. \label{fig:baker1} }
\end{figure}
\begin{figure}
    \centering
    \begin{overpic}[width=14cm]{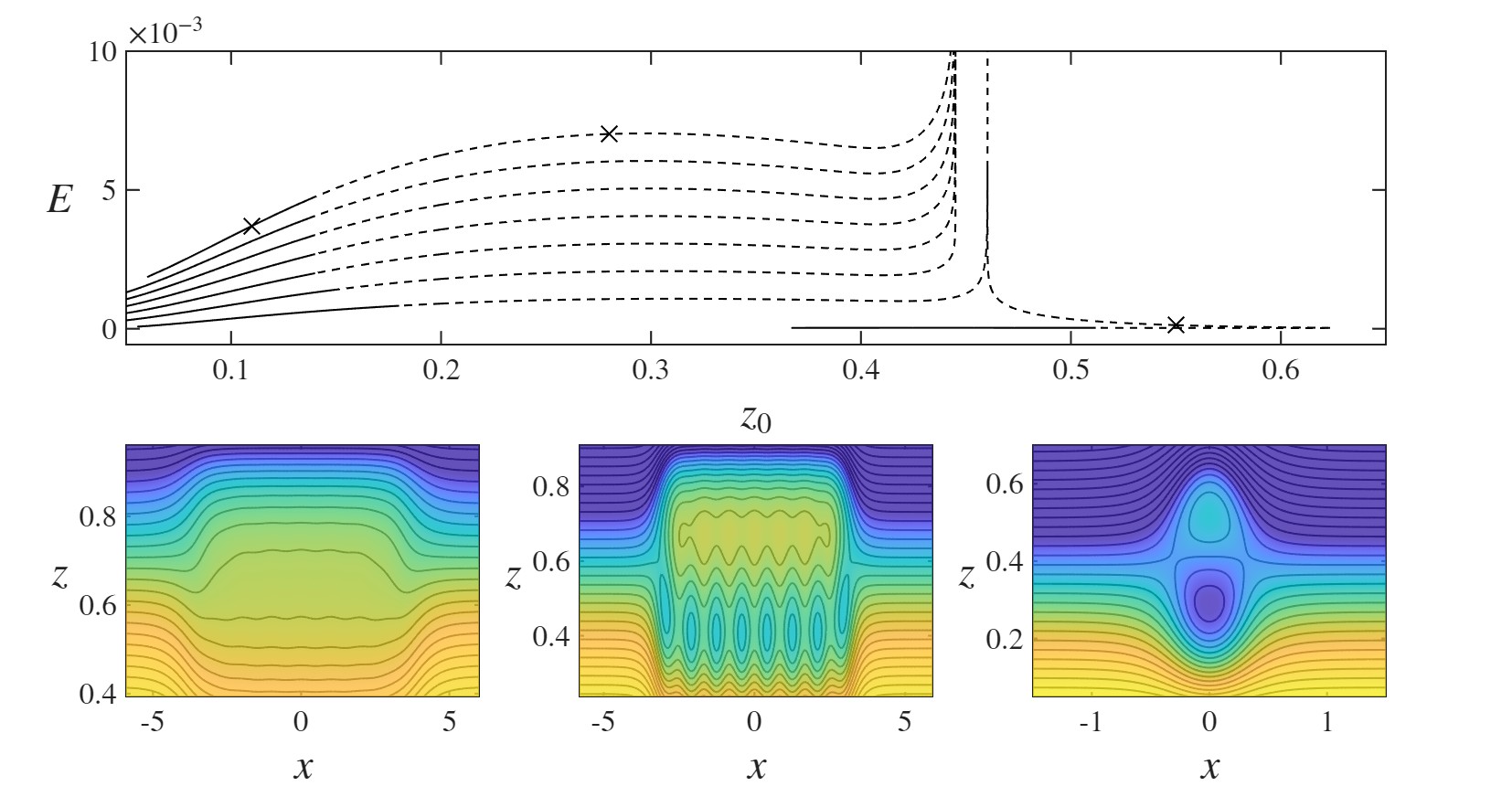}
    \put(2,21.5){$(a)$}
    \put(32,21.5){$(b)$}
    \put(62,21.5){$(c)$}

    \put(16,39){$(a)$}
    \put(38,45){$(b)$}
    \put(78,32){$(c)$}
    \end{overpic}
    \caption{ESW branches for the stratification \eqref{eq:strat_ml}, shown in the upper panel in the $(z_0,E)$-space. For small $z_0$, multihump profiles with no trapped cores, such as that shown in panel $(a)$, are seen. As $z_0$ increases, they develop trapped cores (panel $(b)$). For larger $z_0$, only one  single-hump branch was found (panel $(c)$).  \label{fig:baker2} }
\end{figure}

Unlike for the double-pycnocline stratification, there are no mode-2 computations for a simpler model to indicate where ESWs might be found. We therefore sweep through branches of GSW solutions for $z_0=0.2$, as shown in figure \ref{fig:baker1}. The upper-left panel shows the solution branches in the $(c/c_0^{(2)},K)$-space. Solutions with $K\neq0$ have non-zero tails, that is, they are GSWs. Starting close to the mode-2 long-wave speed, we see a sequence of ``u-shaped'' and ``n-shaped'' branches, in the terminology of \cite{champneys2002true}. At $c\approx1.2c_0^{(2)}$, we find a branch that passes through $K=0$ several times. This branch is seen more clearly in the upper-right panel, where we plot it in the $(E,K)$-space. Several ESWs along this branch are shown in panels $(a)$--$(f)$. The ESWs again have a multihump structure, with each successive ESW gaining an extra oscillation along the broadened section of the wave. All the solutions in figure \ref{fig:baker1} have trapped cores, with an intricate structure not observed for mode-1 ISWs.

Having obtained isolated ESWs along a GSW branch, we trace branches of ESWs by continuation in $z_0$ with $K=0$ imposed. This is shown in figure \ref{fig:baker2}, where the upper panel shows the ESW branches in $(z_0,E)$-space. For small $z_0$ the solutions have no trapped cores ({\it e.g.}\ solution $(a)$), while trapped cores develop as $z_0$ increases (solution $(b)$). As $z_0$ increases, the branches asymptote to a fixed value of $z_0$, corresponding to a broadened conjugate state with oscillations. The branches have a multihump structure: at fixed $z_0$, each successive branch (in order of increasing $E$) has an additional oscillation on the central pulse. Repeating the GSW search of figure \ref{fig:baker1} for $z_0=0.5$ and $z_0=0.8$ revealed only one additional branch, on which solution $(c)$ lies. This solution has a more typical single-hump structure.

\section{Conclusion}\label{sec:conclusion}
In this paper, we provide the first numerical computations of mode-2 solitary wave solutions to the DJL equation in stratifications that are not symmetric about mid-depth. Rather than working with a constrained energy functional, as in \cite{turkington1991computational}, we work directly with the Euler--Lagrange equation. This allows us to find solutions that are saddles, rather than minimisers, of the functional. We demonstrate that mode-2 ESWs occur in  double-pycnocline stratifications, as encountered in laboratory experiments and field studies, and in a surface mixed-layer stratification typical of the upper ocean. We hope this numerical evidence will motivate further analytical work towards establishing their existence rigorously.

A significant outcome of this study is that such mode-2 solutions can now be explored in time-dependent simulations and in laboratory experiments. The key open questions concern the stability of these waves and how they might be generated. One could expect the convex mode-2 ESWs to be unstable, since perturbations generally excite a mode-1 resonant tail. This tail has the same phase speed as the leading solitary wave, but a lower group velocity, and hence continually drains energy from the leading wave. In \cite{carr2015experiments}, convex mode-2 waves were generated by lock release, and these resonant tails are indeed visible, becoming more pronounced as the pycnoclines are moved further away from the channel centre. However, the lock parameters (the difference in depths behind the lock gate) have not yet been explored extensively.  The numerical evidence for the existence of such steady ESWs suggests that it may be possible to generate a mode-2 solitary wave with small tails, which would be longer-lived than those with larger tails. The effect of this resonance on stability, compared with other sources of energy loss ({\it e.g.}\ viscosity), is the subject of ongoing work.

The concave solitary waves with $c>c_0^{(1)}$, on the other hand, do not suffer from this resonance, since they travel faster than all linear mode-1 waves. Hence, it may be possible to generate these waves by lock release so that the resulting wave is steady. The question then becomes which wave is generated: we have shown that there are a large number of coexisting solutions for a given stratification. This multiplicity of solutions may have an interesting effect on the resulting dynamics.

\backsection[Data availability statement]{The code used to generate the results in this study is openly available on GitHub at \url{https://github.com/alexdoak/DJL_FD}.}

\backsection[Funding]{A.D. would like to acknowledge funding from an EPSRC NFFDy Fellowship (EPSRC grant EP/X028607/1). R.B., A.D. and P.A.M. would like to acknowledge funding from Leverhulme Trust (RPG-2023-264). }

\bibliographystyle{abbrvnat}
\bibliography{DJL}
\end{document}